# Robust Transceiver Design Based on Interference Alignment for Multi-User Multi-Cell MIMO Networks with Channel Uncertainty

XIANZHONG XIE[1], HELIN YANG[12], (Student Member, IEEE) AND ATHANASIOS V. VASILAKOS[3], (Senior Member, IEEE)

[1]Chongqing Key Lab of Computer Network and Communication Technology, Chongqing University of Posts and Telecommunications, Chongqing, P. R. China
[2]School of Electrical and Electronic Engineering, Nanyang Technological University, Singapore
[3]Department of Computer and Telecommunications Engineering, University of Western Macedonia, Greece

Corresponding author: H. L. Yang (HYANG013@e.ntu. edu.sg or yhelincqupt@163.com)

**ABSTRACT** In this paper, we firstly exploit the inter-user interference (IUI) and inter-cell interference (ICI) as useful references to develop a robust transceiver design based on interference alignment for a downlink multi-user multi-cell multiple-input multiple-output (MIMO) interference network under channel estimation error. At transmitters, we propose a two-tier transmit beamforming strategy, we first achieve the inner beamforming direction and allocated power by minimizing the interference leakage as well as maximizing the system energy efficiency, respectively. Then, for the outer beamformer design, we develop an efficient conjugate gradient Grassmann manifold subspace tracking algorithm to minimize the distances between the subspace spanned by interference and the interference subspace in the time varying channel. At receivers, we propose a practical interference alignment based on fast and robust fast data projection method (FDPM) subspace tracking algorithm, to achieve the receive beamformer under channel uncertainty. Numerical results show that our proposed robust transceiver design achieves better performance compared with some existing methods in terms of the sum rate and the energy efficiency.

## I. INTRODUCTION

The fifth-generation (5G) technology is a potential technology to improve networks performance for amount of mobile users in wireless multi-cell networks. However, the improvement of the capacity and energy efficiency in wireless networks are limited by generated interference. Hence, the literature [1] [2] studied that interference exploitation (advanced interference aware techniques) is potential way to improve the achievable quality, the reliability, as well as the security of wireless networks, instead of mitigating interference. Moreover, it indicated that beamforming based on interference-aware techniques can enhance the sum rate and energy efficiency through exploring such inter-user interference (IUI) and inter-cell interference (ICI) in 5G wireless systems [2] [3].

Beamforming based on interference-aware techniques has recently been formulated to control the interference in various works, e.g. [3]–[7], to improve the performance of networks. In multi-cell networks, an interference aware-coordinated beamforming design in [3] was developed to control both the inter-user interference and inter-cell interference well. Moreover, recent works have concentrated on multiple-input multiple-output (MIMO) technology based on interference alignment in multi-cell multi-user interfering broadcast channels [5]–[7]. A downlink beamforming and pilot contamination precoding transmission scheme was developed in [5] to mitigate the IUI and ICI with cooperation between cells using a non-iterative closed-form approach. The authors in [6] considered the partially channel state information (CSI) situation in MIMO interfering channels when designed interference-aware beamforming model. Moreover, the authors in [7] investigated the energy efficiency (EE) maximization based on interference alignment for MIMO interfering broadcast channels.

It seems that the above beamforming schemes [2]–[7] can mitigate interferences well in multi-user multi-cell MIMO interfering broadcast channels. However, the performance of the above beamforming schemes may be limited in frequency division duplexing (FDD) scenarios, and the CSI feedback overhead for the downlink mobile users in massive FDD multi-user networks can be overwhelming. Therefore, two-stage beamforming schemes for the MIMO networks were proposed in [8]–[12] recently, which mainly have two steps: an outer beamformer as well as an inner beamformer. References [8] and [9] developed two-stage "hybrid" beamforming schemes in MIMO systems and achieved available performance, but these two methods have correspondingly large number of beamforming chains which may be too power consuming or expensive. Chen and Lau in [10] proposed an outer beamformer design with subspace alignments by minimizing the sum inter-group interference power from other groups and minus the weighted all desired group signal power. Nam *et al.* in [11] developed a two-stage downlink opportunistic beamforming through the user grouping in the FDD large-system regime (both antennas as well as users growing large). Moreover, the authors in [12] modeled a two-stage beamforming framework based on signal-to-leakage-plus-noise ratio (SLNR), and transformed the problem as a trace quotient problem, which aims to enhance significant sum rate gain compared with some existing algorithms. However, it is necessary to note that the above schemes [2]–[12] assumed the perfect CSI in the systems. Due to the dynamic nature of wireless communication environment in cellular networks, information uncertainties naturally occur so that the assumption of perfect CSI may not be practical [13] in MIMO networks.

For the cases of CSI uncertainty, robust versions of beamforming optimization for interference mitigation and management (robust interference alignment (IA) algorithms) models have been studied in [14]–[21] recently. Masouros et al. in [16] proposed a robust beamforming scheme to deal with imperfect CSI, which also reduces the transmit power compared with conventional techniques, while satisfying the required QoS. In [17], a robust IA scheme has been proposed for MIMO interference channel, where only the interference signals spills outside the interference subspace are minimized by using a semidefinite programming (SDP) with high computational complexity. Then, authors in [18] developed a new robust interference subspace beamformer to minimize the interference leakage as well as the desired signal leakage against channel uncertainties. An outage-based robust beamforming scheme for MIMO interference channels was developed in [19] to enhance the sum rate under imperfect CSI. An optimal assignment beamforming scheme was developed [20] in multi-user interference downlink channel by applying a gradient descent method to migrate both ICI and IUI, where it requires only partial CSI at the transmitter. Guiazon et al. in [21] aimed to derive an achievable capacity lower bound for IA and achieve the degree-of-freedom (DoF) of an interference channel with bounded errors. However, the computational complexity for the above beamfoming algorithms scales quickly with the increase number of transmit antenna so low complexity beamforming schemes are needed to be developed for MIMO systems. Moreover, these two-stage beamforming schemes or beamforming based on IA frameworks aim to design transmit or receive beamforming vectors to migrate the interference at the mobile users' receivers without taking into account energy efficiency (or greenness), which is a growing key issue in the telecommunication industry [22]. Most two-stage beamforming schemes aimed at the minimization of interference without subject to suppressing interference to satisfy minimum SLNR constraints, when design the inner or outer beamformer.

In this paper, we firstly design a robust transceiver for multi-user multi-cell MIMO interference networks with channel uncertainties by applying the interference alignment and subspace tracking algorithms. For clarity, the main novelty and contributions of our paper can be summarized:
1) Differing from [10] and [12], this paper exploit interference to formulate a robust two-tier transmit and receive beamforming scheme for a downlink multi-user multi-cell MIMO interference network with imperfect CSI, which designs effective beamformer and power allocation to improve the achievable rate and energy efficiency.
2) We develop a novel two-tier transmit beamformer strategy to achieve inner and outer beaformaning matrices. For the inner beamformer, the beamforming direction and energy-efficient allocated power allocation is achieved by minimizing the IUI. Then, we design a modified set-membership conjugate gradient Grassmann manifold subspace tracking algorithm to achieve the outer beamforming matrix by minimizing ICI in the time varying MIMO interference channels.
3) For the negative effect of channel error on the design of receive beamforming matrix, we present a practical interference alignment based on fast and stable data projection method (FDPM) subspace tracking algorithms, to achieve the receive beamforming vector effectively with a simple change of the step size parameter. The robust minor subspace tracking achieves the orthonormality of the tracked subspace.

The rest of this paper is organized as follows. In Section II, we describe the system model, and formulate the problem optimization framework. Section III presents the robust solutions to achieve the inner beamforming direction, energy-efficient allocated power, and out beamformer under the channel uncertainty model. In Section IV, we derive an efficient IA based on stable subspace tracking algorithm to achieve receive beamformer. The numerical simulations analysis is shown in the Section V. Finally, Section VI concludes the paper.

The following notations are applied in the paper. $(\cdot)^T$ and $(\cdot)^H$ denotes the matrix transpose and conjugate transpose, $\mathbf{I}_{M\times N}$ denotes an $M \times N$ identity matrix; $E\{\cdot\}$ means the expectation operator. $\text{Tr}(\mathbf{A})$ is the trace of a square matrix $\mathbf{A}$. $\|\mathbf{A}\|$ means the Frobenius norm, respectively. Symbol $\nabla f(\cdot)$ denotes the gradient process of the function.

II. SYSTEM MODEL AND PROBLEM FORMULATION

A. System Model

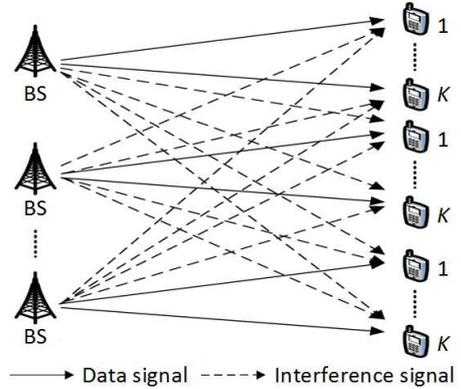

FIGURE 1. System model.

Consider a cellular downlink multi-user multi-cell MIMO network consisting of $B$ cells with $K$ mobile users being in each cell, shown in Fig 1, where each cell only has a base station (BS), and each BS is equipped with $M$ transmission antennas, and each mobile user has $N$ receive antennas. Note that, we consider that the $K$ mobile users are all at the edge boundary of cells, and they will suffer the ICI from other adjacent cells. The downlink channel between the $b'$ th BS to the user $k$ in the $b$ th cell is denoted by $\mathbf{H}_{kb}^{b'} \in \mathbb{C}^{N \times M}$. We set the inner beamforming matrices for the mobile user $k$ in cell $b$ as $\mathbf{V}_{kb} \in \mathbb{C}^{m_b \times d_{kb}}$, where $d_{kb}$ denote the number of data streams transmit for the mobile user $k$ in cell $b$, and the parameter $m_b$ decides the subspace dimension for inter-user spatial multiplexing. The outer beamforming matrix for the cell $b$ is $\mathbf{F}_b \in \mathbb{C}^{M \times m_b}$, $\mathbf{U}_{kb}$ denotes the receiver shaping matrix to the received signal at the $k$ th receiver. Then, the $k$ th user in the cell $b$ linearly process their own received signal to obtain

$$\mathbf{y}_{kb} = \underbrace{\mathbf{U}_{kb}^H \mathbf{H}_{kb}^b \mathbf{F}_b \mathbf{V}_{kb} \mathbf{s}_{kb}}_{\text{desired signal}} + \underbrace{\mathbf{U}_{kb}^H \mathbf{H}_{kb}^b \mathbf{F}_b \sum_{j=1, j\neq k}^{K} \mathbf{V}_{jb} \mathbf{s}_{jb}}_{\text{inter-user interference}} \\ + \underbrace{\mathbf{U}_{kb}^H \mathbf{H}_{kb}^{b'} \mathbf{F}_{b'} \sum_{b' \neq b}^{B} \sum_{i=1}^{K} \mathbf{V}_{ib'} \mathbf{s}_{ib'}}_{\text{inter-cell interference}} + \mathbf{U}_{kb}^H \mathbf{n}_{kb} \quad (1)$$

where $\mathbf{s}_{kb} \in \mathbb{C}^{d_{kb}}$ is the symbol transmitted at the $b$ th BS to the user $k$, the second and third terms in (1) are the inter-user and inter-cell interference, respectively. $\mathbf{n}_{kb} \sim (0, \delta^2 \mathbf{I}_{d_k})$ denote the additive complex Gaussian noise with the noise power $\delta^2$. Note that the beamforming matrix should satisfy a maximum transmission power constraint in the BS given by $\sum_{k=1}^{K} \text{Tr}(\mathbf{F}_b \mathbf{V}_{kb} \mathbf{V}_{kb}^H \mathbf{F}_b^H) = \sum_{k=1}^{K} \text{Tr}(\mathbf{V}_{kb} \mathbf{V}_{kb}^H) \leq P_T$ when we assume that $\mathbf{F}_b^H \mathbf{F}_b = \mathbf{I}_{m_b}$.

In the time-varying channel model, the small time scale channel variation $\mathbf{H}_{kb}^{b'}$ can be applied by the widely used Gauss–Markov channel model [23], given by

$$\mathbf{H}_{kb}^{b'}(t) = \alpha \mathbf{H}_{kb}^{b'}(t-1) + \sqrt{1-\alpha^2} \mathbf{G}_{kb}^{b'}(t) \quad (2)$$

where $\alpha = J_0(2\pi\Omega f_d)$ is the coefficient model of the channel coefficient between the transmitter at the BS $b$ and the user $k$, $J_0(\cdot)$ is the zero-order Bessel function of the first kind. $f_d$ is the maximum Doppler frequency shift, and $\Omega$ is the symbol period. $\text{vec}(\mathbf{G}_{kb}^{b'}(t)) \sim (0, \mathbf{I}_{M \times N})$, $\mathbf{G}_{kb}^{b'}(t)$ and $\mathbf{H}_{kb}^{b'}(0)$ are both independent.

The received signal-to-leakage-plus-noise ratio (SLNR) at the user $k$ in the $b$ th cell can be given as

$$\gamma_{kb} = \frac{\|\mathbf{H}_{kb}^b \mathbf{F}_b \mathbf{V}_{kb}\|^2}{\sum_{j \neq k}^{K} \|\mathbf{H}_{jb}^b \mathbf{F}_b \mathbf{V}_{kb}\|^2 + \sum_{b' \neq b}^{B} \sum_{i=1}^{K} \|\mathbf{H}_{ib'}^b \mathbf{F}_b \mathbf{V}_{kb}\|^2 + \delta^2} \quad (3)$$

In real scenarios, imperfect CSI should be taken into account which brings channel estimation error into the system. For the $k$ th user in the cell $b$, to model estimation inaccuracies, e.g., the underlying channel uncertainty $\Delta \mathbf{H}_{kb}^b$. Then, the channel matrix $\mathbf{H}_{kb}^b$ and norm-bounded channel uncertainty $\Delta \mathbf{H}_{kb}^b$ can be given as

$$\mathbf{H}_{kb}^b = \widehat{\mathbf{H}}_{kb}^b + \Delta \mathbf{H}_{kb}^b \\ \mathcal{G}_{kb}^b = \{\Delta \mathbf{H}_{kb}^b | \text{Tr}\{\Delta \mathbf{H}_{kb}^b \Delta \mathbf{H}_{kb}^{b\,H}\} \leq \varepsilon_{kb}\} \quad (4)$$

where $\widehat{\mathbf{H}}_{kb}^b$ is the estimated channel, which can be known at the BS transmitter, and $\varepsilon_{kb}$ is the radius of the channel uncertainty region. The above norm-bounded uncertainty model results in the worst-case situation that ensure the corresponding interference migration scheme is protected for each realization of the channel uncertain portion of the interference channels. Hence, the performance of the robust algorithm based on this uncertainty model will provides benchmark performance.

Note that $\Delta \mathbf{H}_{kb}^b$ is not correlated with the accurate channel matrix $\mathbf{H}_{kb}^b$. In practice, the radius as well as the shape of the above uncertainty region depends on the second-order estimated channel error statistics. If $\Delta \mathbf{H}_{kb}^b$ is assumed zero mean and covariance matrix $\text{E}\{\Delta \mathbf{H}_{kb}^b \Delta \mathbf{H}_{kb}^{b\,H}\} = \delta_e^2 \mathbf{I}_{d_k}$, the radius of the channel uncertainty is modeled as $\varepsilon_{kb}^2 = \tau \delta_e^2$, where $\tau > 0$ [24], For simplicity, we make equal channel estimation error variance assumption $\delta_{1b}^2 = ,\cdots, = \delta_{Kb}^2 = \delta_{1b'}^2, \cdots, \delta_{KB}^2 = \delta_e^2$. If $\text{E}\{\Delta \mathbf{H}_{kb}^b \Delta \mathbf{H}_{kb}^{b\,H}\} \neq \delta_e^2 \mathbf{I}_{d_k}$, Then, the channel uncertainty region is also set as the ellipsoid uncertainty region, such as $\mathcal{G}_{kb}^b = \{\Delta \mathbf{H}_{kb}^b | \text{Tr}\{\Delta \mathbf{H}_{kb}^b (\text{E}(\Delta \mathbf{H}_{kb}^b \Delta \mathbf{H}_{kb}^{b\,H}))^{-1} \Delta \mathbf{H}_{kb}^{b\,H}\} \leq \varepsilon_{kb}\}$ [24].

Based on the error model (4), the robust SLNR of the $k$ th user becomes a function of the estimated channel and the random errors, which is expressed

$$\widehat{\gamma}_{kb} = \frac{\|(\widehat{\mathbf{H}}_{kb}^b + \Delta \mathbf{H}_{kb}^b) \mathbf{F}_b \mathbf{V}_{kb}\|^2}{\sum_{j \neq k}^{K} \|(\widehat{\mathbf{H}}_{jb}^b + \Delta \mathbf{H}_{jb}^b) \mathbf{F}_b \mathbf{V}_{kb}\|^2 + \sum_{b' \neq b}^{B} \sum_{i=1}^{K} \|(\widehat{\mathbf{H}}_{ib'}^b + \Delta \mathbf{H}_{ib'}^b) \mathbf{F}_{b'} \mathbf{V}_{kb}\|^2 + \delta^2}$$

(5)

### B. Problem Formulation

In order to mitigate the cross-tier and co-tier interferences from other cells in the transmissions, before developing the optimization problem, we give the definition of interference leakage firstly. Each cell $b$ computes the total leakage of interference (LIF) to other unserving users or cells showing the amount of generated or misaligned interference when it transmits data streams to the user $k$ in its cell $b$, which is expressed by

$$LIF_{kb} = \underbrace{\sum_{j \neq k}^{K} \|(\widehat{\mathbf{H}}_{jb}^b + \Delta \mathbf{H}_{jb}^b) \mathbf{F}_b \mathbf{V}_{kb}\|^2}_{\text{inter-user interference leakage}} + \underbrace{\sum_{b' \neq b}^{B} \sum_{i=1}^{K} \|(\widehat{\mathbf{H}}_{ib'}^{b'} + \Delta \mathbf{H}_{ib'}^{b'}) \mathbf{F}_b \mathbf{V}_{kb}\|^2}_{\text{inter-cell interference leakage}} \quad (6)$$

From (6), the transmit beamforming can be considered to minimize the total leakage of interference of each cell while maintaining the required SLNR of each user at a certain level. In order to achieve both the inner beamforming vector $\mathbf{V}_{kb}$ and outer beamforming matrix $\mathbf{F}_b$, we can divide the LIF minimization problem into two sub-problems (**P1.1** and **P1.2**). For the sub-problems **P1.1**, we can obtain $\mathbf{V}_{kb}$ by minimizing the inter-user interference leakage when $\mathbf{F}_b$ is fixed. For the sub-problems **P1.2**, we can get $\mathbf{F}_b$ by minimizing the inter-cell interference leakage when $\mathbf{V}_{kb}$ is fixed.

Then, the sub-problems **P1.1** can be formulated as

$$\textbf{P1.1} \min_{\mathbf{V}_{1b}, \cdots, \mathbf{V}_{Kb}} \sum_{k=1}^{K} (\widehat{\mathbf{H}}_{jb}^b + \Delta \mathbf{H}_{jb}^b) \mathbf{F}_b \mathbf{V}_{kb} \\ s.t. \ \widehat{\gamma}_{kb} \geq \overline{\gamma}_{kb}, \quad k = 1, \cdots, K \quad (7)$$

where $\overline{\gamma}_{kb}$ denotes the minimum SLNR requirement of the $k$ th user in the cell $b$.

In addition, the sub-problems **P1.2** is expressed as

$$\textbf{P1.2} \min_{\mathbf{F}_b} \sum_{k=1}^{K} \left( \sum_{b' \neq b}^{B} \sum_{i=1}^{K} \|(\widehat{\mathbf{H}}_{ib'}^b + \Delta \mathbf{H}_{ib'}^b) \mathbf{F}_b \mathbf{V}_{kb}\|^2 \right) \\ s.t. \ \mathbf{F}_b^H \mathbf{F}_b = \mathbf{I}_{m_b} \quad (8)$$

For the $k$ th user in the cell $b$, applying the receiver shaping matrix $\mathbf{U}_{kb}^H$, the total undesired interference signal received from the inter-user interference from other users in the same cell $b$ and inter-cell interference from the other cells as follows

$$I_{kb} = \sum_{j \neq k}^{K} \left\| \mathbf{U}_{kb}^{H} (\hat{\mathbf{H}}_{jb}^{b} + \Delta \mathbf{H}_{jb}^{b}) \mathbf{F}_{b} \mathbf{V}_{jb} \right\|^{2} + \sum_{b' \neq b}^{B} \sum_{i=1}^{K} \left\| (\mathbf{U}_{kb}^{H} \mathbf{H}_{kb}^{b'} + \Delta \mathbf{H}_{kb}^{b'}) \mathbf{F}_{b'} \mathbf{V}_{ib'} \right\|^{2}$$
(9)

Then, the robust interference minimization problem for the mobile user $k$ in the cell $b$ can be expressed as

$$\mathbf{P2} \quad \min_{\mathbf{U}_{kb}} \quad I_{kb}$$
$$\text{subject to} \quad \mathbf{U}_{kb}^{H} \mathbf{U}_{kb} = \mathbf{I}_{d_b}$$
(10)

The next two sections will present robust solutions to achieve two-tier transmit beamformer (details shown in section III by solving **P1.1** and **P1.2** and receive beamformer (details shown in section IV by solving **P2**).

### III. ROBUST SOLUTIUONS FOR TWO-TIER TRANSMIT BEAMFORMER

In this section, we will design robust schemes to achieve both inner and outer beamformers under channel errors, First, given the estimated channel $\hat{\mathbf{H}}_{kb}^{b}$ at transmitter, the objective function in **P1** is obtained by the random error into account, then, the uncertain values in **P1** can be achieved by

$$\left\| (\hat{\mathbf{H}}_{kb}^{b} + \Delta \mathbf{H}_{kb}^{b}) \mathbf{F}_{b} \mathbf{V}_{kb} \right\|^{2}$$
$$= \mathbf{V}_{kb}^{H} \mathbf{F}_{b}^{H} (\hat{\mathbf{H}}_{kb}^{b} + \Delta \mathbf{H}_{kb}^{b})^{H} (\hat{\mathbf{H}}_{kb}^{b} + \Delta \mathbf{H}_{kb}^{b}) \mathbf{F}_{b} \mathbf{V}_{kb}$$
$$\stackrel{(a)}{=} \text{Tr} \left\{ \left( \hat{\mathbf{H}}_{kb}^{b\,H} \hat{\mathbf{H}}_{kb} + \Delta \mathbf{H}_{kb}^{b\,H} \Delta \mathbf{H}_{kb} \right) \mathbf{F}_{b} \mathbf{V}_{kb} \mathbf{V}_{kb}^{H} \mathbf{F}_{b}^{H} \right\}$$
$$= \text{Tr} \left\{ \left( \hat{\mathbf{H}}_{kb}^{b\,H} \hat{\mathbf{H}}_{kb}^{b} + \delta_{e}^{2} \mathbf{I}_{k} \right) \mathbf{F}_{b} \mathbf{V}_{kb} \mathbf{V}_{kb}^{H} \mathbf{F}_{b}^{H} \right\}$$
(11)

where the step (a) can be got from the independence between the estimated channel $\hat{\mathbf{H}}_{kb}^{b}$ and $\Delta \mathbf{H}_{kb}^{b}$, and consequently yielding $\hat{\mathbf{H}}_{kb}^{b\,H} \Delta \mathbf{H}_{kb}^{b} = 0$ and $\hat{\mathbf{H}}_{kb}^{b} \Delta \mathbf{H}_{kb}^{b\,H} = 0$.

Similarly, we can get the following equations

$$\sum_{j \neq k}^{K} \left\| (\hat{\mathbf{H}}_{jb}^{b} + \Delta \mathbf{H}_{jb}^{b}) \mathbf{F}_{b} \mathbf{V}_{kb} \right\|^{2} = \sum_{j \neq k}^{K} \text{Tr} \left\{ \left( \hat{\mathbf{H}}_{jb}^{b\,H} \hat{\mathbf{H}}_{jb}^{b} + \delta_{e}^{2} \mathbf{I}_{d_k} \right) \mathbf{F}_{b} \mathbf{V}_{kb} \mathbf{V}_{kb}^{H} \mathbf{F}_{b}^{H} \right\}$$
$$\sum_{b' \neq b}^{B} \sum_{i=1}^{K} \left\| (\hat{\mathbf{H}}_{ib'}^{b} + \Delta \mathbf{H}_{ib'}^{b}) \mathbf{F}_{b} \mathbf{V}_{kb} \right\|^{2} = \sum_{b' \neq b}^{B} \sum_{i=1}^{K} \text{Tr} \left\{ \left( \hat{\mathbf{H}}_{ib'}^{b\,H} \hat{\mathbf{H}}_{ib'}^{b} + \delta_{e}^{2} \mathbf{I}_{d_k} \right) \mathbf{F}_{b} \mathbf{V}_{kb} \mathbf{V}_{kb}^{H} \mathbf{F}_{b}^{H} \right\}$$
(12)

For the above analysis for channel error model, the robust LIF minimization two sub-problems **P1.1** and **P1.2** can be expressed as

$$\mathbf{P1.1} \quad \min_{\mathbf{V}_{1b}, \cdots, \mathbf{V}_{Kb}} \sum_{k=1}^{K} \left( \sum_{j \neq k}^{K} \text{Tr} \left\{ \left( \hat{\mathbf{H}}_{jb}^{b\,H} \hat{\mathbf{H}}_{jb}^{b} + \delta_{e}^{2} \mathbf{I}_{d_k} \right) \mathbf{F}_{b} \mathbf{V}_{kb} \mathbf{V}_{kb}^{H} \mathbf{F}_{b}^{H} \right\} \right)$$
$$s.t. \quad \gamma_{kb} \geq \overline{\gamma}_{kb}, \quad k = 1, \cdots, K$$

$$\mathbf{P1.2} \quad \min_{\mathbf{F}_{b}} \sum_{k=1}^{K} \left( \sum_{b' \neq b}^{B} \sum_{i=1}^{K} \text{Tr} \left\{ \left( \hat{\mathbf{H}}_{ib'}^{b\,H} \hat{\mathbf{H}}_{ib'}^{b} + \delta_{e}^{2} \mathbf{I}_{d_k} \right) \mathbf{F}_{b} \mathbf{V}_{kb} \mathbf{V}_{kb}^{H} \mathbf{F}_{b}^{H} \right\} \right)$$
$$s.t. \quad \mathbf{F}_{b}^{H} \mathbf{F}_{b} = \mathbf{I}_{m_b}$$
(13)

From (13), the inner beamforming vectors $\mathbf{V}_{kb}$ can be achieved through solving **P1.1** with only needs the local CSI from the BS and its serving mobile users when $\mathbf{F}_b$ is fixed; the outer beamforming matrix $\mathbf{F}_b$ can be obtained by solving **P1.2** which only requires the channel statistic cross cells when $\mathbf{V}_{kb}$ is fixed. Finally, we can obtain the optimal $\mathbf{V}_{kb}$ and $\mathbf{F}_b$ through applying a distributed iterative algorithm.

#### A. Inner Beamformer Design with SLNR constraints

Without loss of generality, we aim to reformulate **P1.1** as a convex problem. The objective function in **P1.1** is obviously a convex function for the beamforming vectors $\mathbf{V}_{kb}$. A trick from is applied to get the hidden convexity of SLNR constraints, $\gamma_{kb} \geq \overline{\gamma}_{kb}$. We note that all the values (such as $\|\cdot\|^2$) in the SLNRs in (5) are real valued and positive. the constraint SLNR $\gamma_{kb} \geq \overline{\gamma}_{kb}$ can be rewritten as (14), as shown at the top of the next page.

Clearly, the reformulated SLNR constraint in (13) is a second-order constraint, which is a convex constraint [25]. Hence, optimization theory can provide some important properties for the above reformulated convex problem; we apply Karush–Kuhn–Tucker (KKT) conditions and Lagrange dual decomposition method (LDDM) optimization to solve the reformulated convex problem. Hence, the corresponding Lagrangian function of **P1.1** can be expressed as (15), where $\lambda_{kb}$ is the non-negative Lagrangian multiplier for the $k$ th user SLNR constraint in the cell $b$. The dual function is $[\mathbf{V}_{1b}, \cdots, \mathbf{V}_{Kb}] = \arg \min \sum_{k=1}^{K} \lambda_k$, and the duality implies equals the total LIF at the optimal solution.

Then, using the following definition for differentiation of $\mathcal{L}_{kb}(\mathbf{V}_{1b}, \cdots, \mathbf{V}_{Kb}, \lambda_{1b}, \cdots, \lambda_{Kb})$ with respect to $\mathbf{V}_{kb}$,

$$\frac{\text{Tr} \left\{ \left( \hat{\mathbf{H}}_{kb}^{b\,H} \hat{\mathbf{H}}_{kb}^{b} + \delta_{e}^{2} \mathbf{I}_{d_k} \right) \mathbf{F}_{b} \mathbf{V}_{kb} \mathbf{V}_{kb}^{H} \mathbf{F}_{b}^{H} \right\}}{\delta^{2} \overline{\gamma}_{kb}} \geq \sqrt{\frac{1}{\delta^{2}} \left( \sum_{j \neq k}^{K} \text{Tr} \left\{ \left( \hat{\mathbf{H}}_{jb}^{b\,H} \hat{\mathbf{H}}_{jb}^{b} + \delta_{e}^{2} \mathbf{I}_{d_k} \right) \mathbf{F}_{b} \mathbf{V}_{kb} \mathbf{V}_{kb}^{H} \mathbf{F}_{b}^{H} \right\} + \sum_{b' \neq b}^{B} \sum_{i=1}^{K} \text{Tr} \left\{ \left( \hat{\mathbf{H}}_{ib'}^{b\,H} \hat{\mathbf{H}}_{ib'}^{b} + \delta_{e}^{2} \mathbf{I}_{d_k} \right) \mathbf{F}_{b} \mathbf{V}_{kb} \mathbf{V}_{kb}^{H} \mathbf{F}_{b}^{H} \right\} \right) + 1}$$
(14)

$$\mathcal{L}_{b}(\mathbf{V}_{1b}, \cdots, \mathbf{V}_{Kb}, \lambda_{1b}, \cdots, \lambda_{Kb}) = \sum_{k=1}^{K} \left( \sum_{j \neq k}^{K} \text{Tr} \left\{ \left( \hat{\mathbf{H}}_{jb}^{b\,H} \hat{\mathbf{H}}_{jb}^{b} + \delta_{e}^{2} \mathbf{I}_{d_k} \right) \mathbf{F}_{b} \mathbf{V}_{kb} \mathbf{V}_{kb}^{H} \mathbf{F}_{b}^{H} \right\} \right) + \sum_{k=1}^{K} \lambda_{kb} \left( \frac{1}{\delta^{2}} \left( \sum_{j \neq k}^{K} \text{Tr} \left\{ \left( \hat{\mathbf{H}}_{jb}^{b\,H} \hat{\mathbf{H}}_{jb}^{b} + \hat{\mathbf{H}}_{ib'}^{b\,H} \hat{\mathbf{H}}_{ib'}^{b} + 2\delta_{e}^{2} \mathbf{I}_{d_k} \right) \mathbf{F}_{b} \mathbf{V}_{kb} \mathbf{V}_{kb}^{H} \mathbf{F}_{b}^{H} \right\} \right)$$
$$+ 1 - \frac{\text{Tr} \left\{ \left( \hat{\mathbf{H}}_{kb}^{b\,H} \hat{\mathbf{H}}_{kb}^{b} + \delta_{e}^{2} \mathbf{I}_{d_k} \right) \mathbf{F}_{b} \mathbf{V}_{kb} \mathbf{V}_{kb}^{H} \mathbf{F}_{b}^{H} \right\}}{\delta^{2} \overline{\gamma}_{kb}} \right)$$
(15)

$$\frac{\partial \mathcal{L}_{kb}(\mathbf{V}_{kb},\lambda_{kb})}{\partial \mathbf{V}_{kb}} = \left( \sum_{j\neq k}^{K} \frac{\delta^2+\lambda_{jb}}{\delta^2}\left(\hat{\mathbf{H}}_{jb}^{b\,H}\hat{\mathbf{H}}_{jb}^{b}+\hat{\mathbf{H}}_{ib'}^{b\,H}\hat{\mathbf{H}}_{ib'}^{b}+2\delta_e^2 \mathbf{I}_{d_k}\right)\mathbf{F}_b\mathbf{F}_b^H \right)$$
$$\times \mathbf{V}_{kb} - \frac{\lambda_{kb}}{\delta^2}\frac{\mathbf{F}_b^H(\hat{\mathbf{H}}_{kb}+\Delta\mathbf{H}_{kb})^H(\hat{\mathbf{H}}_{kb}+\Delta\mathbf{H}_{kb})\mathbf{F}_b\mathbf{V}_{kb}}{\bar{\gamma}_{kb}} \tag{16}$$

Equating $\partial \mathcal{L}_{kb}(\mathbf{V}_{kb},\lambda_{kb})/\partial \mathbf{V}_{kb}$ to zero, we can get that

$$\frac{\lambda_{kb}}{\delta^2}\frac{\mathbf{F}_b^H(\hat{\mathbf{H}}_{kb}+\Delta\mathbf{H}_{kb})^H(\hat{\mathbf{H}}_{kb}+\Delta\mathbf{H}_{kb})\mathbf{F}_b\mathbf{V}_{kb}}{\bar{\gamma}_{kb}}$$
$$= \left( \sum_{j\neq k}^{K} \frac{\delta^2+\lambda_{jb}}{\delta^2}\left(\hat{\mathbf{H}}_{jb}^{b\,H}\hat{\mathbf{H}}_{jb}^{b}+\hat{\mathbf{H}}_{ib'}^{b\,H}\hat{\mathbf{H}}_{ib'}^{b}+2\delta_e^2\mathbf{I}_{d_k}\right)\mathbf{F}_b\mathbf{F}_b^H \right)\mathbf{V}_{kb} \tag{17}$$

Transforming into an eigenvalue problem by multiplying both sides of (17) with $(\mathbf{H}_{kb}^{S1\,H}\mathbf{H}_{kb}^{S1}\mathbf{F}_b\mathbf{F}_b^H + \sum_{j\neq k}^{K}(\delta^2+\lambda_{jb})(\hat{\mathbf{H}}_{jb}^{b\,H}\hat{\mathbf{H}}_{jb}^{b} + \hat{\mathbf{H}}_{ib'}^{b\,H}\hat{\mathbf{H}}_{ib'}^{b} + 2\delta_e^2\mathbf{I}_{d_k})\mathbf{F}_b\mathbf{F}_b^H/\delta^2)^{-1}$, then, it follows that

$$\mathbf{V}_{kb} = \left( \sum_{j\neq k}^{K}\frac{\delta^2+\lambda_{jb}}{\delta^2}\left(\hat{\mathbf{H}}_{jb}^{b\,H}\hat{\mathbf{H}}_{jb}^{b}+\hat{\mathbf{H}}_{ib'}^{b\,H}\hat{\mathbf{H}}_{ib'}^{b}+2\delta_e^2\mathbf{I}_{d_k}\right)\mathbf{F}_b\mathbf{F}_b^H \right)^{-1} \mathbf{F}_b^H(\hat{\mathbf{H}}_{kb}+\Delta\mathbf{H}_{kb})^H$$
$$\times \underbrace{\frac{\lambda_{kb}}{\delta^2\bar{\gamma}_{kb}}(\hat{\mathbf{H}}_{kb}+\Delta\mathbf{H}_{kb})\mathbf{F}_b\mathbf{V}_{kb}}_{=\text{scalar when } d_{kb} \text{ and } N=1} \tag{18}$$

From (18), due to $\lambda_{kb}(\hat{\mathbf{H}}_{kb}+\Delta\mathbf{H}_{kb})\mathbf{F}_b\mathbf{V}_{kb}/\delta^2\bar{\gamma}_{kb}$ is a scalar when the mobile user is equipped with a single receive antenna and the number of the data streams is 1, $\mathbf{V}_{kb}$ should be parallel to $(\sum_{j\neq k}^{K}(\delta^2+\lambda_{jb})(\hat{\mathbf{H}}_{jb}^{b\,H}\hat{\mathbf{H}}_{jb}^{b}+\hat{\mathbf{H}}_{ib'}^{b\,H}\hat{\mathbf{H}}_{ib'}^{b}+2\delta_e^2\mathbf{I}_{d_k})\mathbf{F}_b\mathbf{F}_b^H/\delta^2)^{-1}(\hat{\mathbf{H}}_{kb}+\Delta\mathbf{H}_{kb})^H\mathbf{F}_b^H$ $(\hat{\mathbf{H}}_{kb}+\Delta\mathbf{H}_{kb})^H$ In other words, the optimal beamforming vectors $\mathbf{V}_{1b},\cdots,\mathbf{V}_{Kb}$ are

$$\mathbf{V}_{kb}^* = \sqrt{p_{kb}} \underbrace{\frac{\left( \sum_{j\neq k}^{K}\frac{\delta^2+\lambda_{jb}}{\delta^2}\left(\hat{\mathbf{H}}_{jb}^{b\,H}\hat{\mathbf{H}}_{jb}^{b}+\hat{\mathbf{H}}_{ib'}^{b\,H}\hat{\mathbf{H}}_{ib'}^{b}+2\delta_e^2\mathbf{I}_{d_k}\right)\mathbf{F}_b\mathbf{F}_b^H \right)^{-1}\mathbf{F}_b^H(\hat{\mathbf{H}}_{kb}+\Delta\mathbf{H}_{kb})^H}{\left\| \left( \sum_{j\neq k}^{K}\frac{\delta^2+\lambda_{jb}}{\delta^2}\left(\hat{\mathbf{H}}_{jb}^{b\,H}\hat{\mathbf{H}}_{jb}^{b}+\hat{\mathbf{H}}_{ib'}^{b\,H}\hat{\mathbf{H}}_{ib'}^{b}+2\delta_e^2\mathbf{I}_{d_k}\right)\mathbf{F}_b\mathbf{F}_b^H \right)^{-1}\mathbf{F}_b^H(\hat{\mathbf{H}}_{kb}+\Delta\mathbf{H}_{kb})^H \right\|}}_{\tilde{\mathbf{V}}_{kb}^* = \text{beamforming direction}}$$
$$\tag{19}$$

where $p_{kb}$ is the beamforming power, and $\tilde{\mathbf{V}}_{kb}^*$ denotes the unit-norm optimal beamforming direction for the user $k$ in the cell $b$. The beamforming powers of all users can be achieved through noting that the SLNR constraints in (13) hold with the equality at the optimal solution. This implies $p_{kb}\text{Tr}\{(\hat{\mathbf{H}}_{kb}^{b\,H}\hat{\mathbf{H}}_{kb}^{b}+\delta_e^2\mathbf{I}_{d_k})\mathbf{F}_b\tilde{\mathbf{V}}_{kb}^*\tilde{\mathbf{V}}_{kb}^{*H}\mathbf{F}_b^H\}/\bar{\gamma}_{kb} - \sum_{j\neq k}^{K}p_{kb}\text{Tr}\{(\hat{\mathbf{H}}_{jb}^{b\,H}\hat{\mathbf{H}}_{jb}^{b}+\hat{\mathbf{H}}_{ib'}^{b\,H}\hat{\mathbf{H}}_{ib'}^{b}+2\delta_e^2\mathbf{I}_{d_k})\mathbf{F}_b\tilde{\mathbf{V}}_{kb}^*\tilde{\mathbf{V}}_{kb}^{*H}\mathbf{F}_b^H\} = \delta^2$ for $k=1,\cdots,K$ in all cells. For we have computed the beamforming directions, we can achieve the beamforming power of the user $k$ in the cell $b$ as

$$p_{kb} = \delta^2 \left( \text{Tr}\left\{ \left(\hat{\mathbf{H}}_{kb}^{b\,H}\hat{\mathbf{H}}_{kb}^{b}+\delta_e^2\mathbf{I}_{d_k}\right)\mathbf{F}_b\tilde{\mathbf{V}}_{kb}^*\tilde{\mathbf{V}}_{kb}^{*H}\mathbf{F}_b^H \right\}/\bar{\gamma}_{kb} \right.$$
$$\left. - \sum_{j\neq k}^{K}\text{Tr}\left\{ \left(\hat{\mathbf{H}}_{jb}^{b\,H}\hat{\mathbf{H}}_{jb}^{b}+\hat{\mathbf{H}}_{ib'}^{b\,H}\hat{\mathbf{H}}_{ib'}^{b}+2\delta_e^2\mathbf{I}_{d_k}\right)\mathbf{F}_b\tilde{\mathbf{V}}_{kb}^*\tilde{\mathbf{V}}_{kb}^{*H}\mathbf{F}_b^H \right\} \right)^{-1} \tag{20}$$

Since we combine (19) and (20), we can get optimal beamforming vectors and powers with a function of Lagrange multipliers $\lambda_{1b},\cdots,\lambda_{Kb}$ in each cell $b$. The Lagrange multipliers $\lambda_{kb}$ can be achieved by applying the convex optimization or a fixed-point equations.

### B. Energy-Efficient Power Allocation

For the above subsection, we compute the beamforming power $p_{kb}$ in (20) which minimizes the inter-user interference leakage to other users in the same cell while guarantee the SLNR constraints. However, these are not the energy efficient power value. Energy efficiency issue is an important issue in current wireless communications, and in this subsection, the energy-efficient beamforming power allocation problem in interference minimization based MIMO networks is analyzed, and the power allocation algorithms aiming at maximizing the energy efficiency of the MIMO network are designed.

The energy efficiency of the cell $b$ is defined as the sum rate to the power consumption [7], which is

$$Q_b(\mathbf{P}_b,\mathbf{P}_{-b}) = \frac{R_b(\mathbf{P}_b,\mathbf{P}_{-b})}{P_b^{\text{total}}(\mathbf{P}_b,\mathbf{P}_{-b})}$$
$$= \frac{\sum_{k=1}^{K}\left(\log_2\det\left(\mathbf{I}_N + P_{kb}|(\hat{\mathbf{H}}_{kb}^b+\Delta\mathbf{H}_{kb}^b)\mathbf{F}_b\tilde{\mathbf{V}}_{kb}^*|^2\mathbf{J}_{bk}^{-1}\right)\right)}{\rho\sum_{k=1}^{K}P_{kb} + MP_b^c + P_b^o} \tag{21}$$

where $\mathbf{P}_b=[P_{1b},\cdots,P_{Kb}]$ is the power allocation strategy response of the cell $b$ given the other cells' power allocation strategies $\mathbf{P}_{-b}$. $R_b$ and $P_b^{\text{total}}$ are the sum rate and total power consumption in the cell $b$. $\rho$ denote the reciprocal efficiency of the power amplifier; $P_{kb}^c$ is the circuit power consumed at every transmit antennas consisting of the duplexers, bandpass filters, and other radio-frequency circuits; and $P_b^o$ is the power consumption at the BS $b$. $\mathbf{J}_{kb}^{-1}$ is the interference plus noise covariance matrix of the user $k$ in the cell $b$

$$\mathbf{J}_{kb}^{-1} = \sum_{j\neq k}^{K}P_{jb}|(\hat{\mathbf{H}}_{jb}^b+\Delta\mathbf{H}_{jb}^b)\mathbf{F}_b\tilde{\mathbf{V}}_{jb}^*|^2 + \sum_{b'\neq b}^{B}\sum_{i=1}^{K}P_{ib'}|(\hat{\mathbf{H}}_{ib'}^{b'}+\Delta\mathbf{H}_{ib'}^{b'})\mathbf{F}_{b'}\tilde{\mathbf{V}}_{ib'}^*|^2 + \mathbf{I}_N\delta^2.$$

The optimization problem in (21) is well-known as NP-hard problems, so it is difficult to directly derive the global optimal solutions and the fractional objective makes it even more complicated. Therefore, we commit to many effective schemes to solve the problem in this section.

The objective function defined in (21) is non-convex, but we can transform it into a concave function by applying the nonlinear fractional programming proposed in [26]. Then, the fractional programming in (21) can be transmitted with the following problem

$$\max R_b(\mathbf{P}_b^*,\mathbf{P}_{-b}) - Q_b P_b^{\text{total}}(\mathbf{P}_b^*,\mathbf{P}_{-b})$$
$$s.t. \ \hat{\gamma}_{kb} \geq \bar{\gamma}_{kb}, \ \forall k=1, \text{ or } P_{kb} \geq p_{kb} \tag{22}$$

where $p_{kb}$ is the beamforming power by minimizing the inference leakage under SLNR constraint in (20), $Q_b^*$ is the maximum EE of (21) with the optimal power allocation strategy $\mathbf{P}_b^*$. Then, we can obtain the following equivalence

$$Q_b = Q_b^* \Leftrightarrow R_b(\mathbf{P}_b^*,\mathbf{P}_{-b}) - Q_b^* P_b^{\text{total}}(\mathbf{P}_b^*,\mathbf{P}_{-b}) = 0 \tag{23}$$

**Theorem 1**: The maximum EE $Q_b^*$ is achieved if and only if

$$\max R_b(\mathbf{P}_b^*, \mathbf{P}_{-b}) - Q_b^* P_b^{\text{total}}(\mathbf{P}_b^*, \mathbf{P}_{-b}) = R_b(\mathbf{P}_b^*, \mathbf{P}_{-b}) - Q_b^* P_b^{\text{total}}(\mathbf{P}_b^*, \mathbf{P}_{-b}) = 0$$

for $R_b(\mathbf{P}_b^*, \mathbf{P}_{-b}) \geq 0$ and $P_b^{\text{total}}(\mathbf{P}_b^*, \mathbf{P}_{-b}) > 0$.

**Proof**: The optimal EE $Q_b^*$ can be expressed as

$$Q_b^* = \frac{R_b(\mathbf{P}_b^*, \mathbf{P}_{-b})}{P_b^{\text{total}}(\mathbf{P}_b^*, \mathbf{P}_{-b})} \geq \frac{R_b(\mathbf{P}_b, \mathbf{P}_{-b})}{P_b^{\text{total}}(\mathbf{P}_b, \mathbf{P}_{-b})} \Rightarrow R_b(\mathbf{P}_b, \mathbf{P}_{-b}) - Q_b^* P_b^{\text{total}}(\mathbf{P}_b, \mathbf{P}_b) \leq 0$$

$$\& R_b(\mathbf{P}_b^*, \mathbf{P}_{-b}) - Q_b^* P_b^{\text{total}}(\mathbf{P}_b^*, \mathbf{P}_{-b}) = 0.$$

Hence, we conclude that $\max R_b(\mathbf{P}_b, \mathbf{P}_{-b}) - Q_b^* P_b^{\text{total}}(\mathbf{P}_b, \mathbf{P}_{-b}) = 0$, and it is obtained by giving the UEs' strategies $\{\mathbf{P}_b^*, \mathbf{P}_{-b}^*\}$.

**Theorem 2**: The convergence implication of the Theorem 1.
**Proof**: For any feasible power control policies $\{\mathbf{P}_b^*, \mathbf{P}_{-b}\}$, we can achieve the following inequality $R_b(\mathbf{P}_b, \mathbf{P}_{-b}) - Q_b^* P_b^{\text{total}}(\mathbf{P}_b, \mathbf{P}_{-b}) \leq R_b(\mathbf{P}_b^*, \mathbf{P}_{-b}) - Q_b^* P_b^{\text{total}}(\mathbf{P}_b^*, \mathbf{P}_{-b}) = 0$.

The preceding inequality implies

$$\frac{R_b(\mathbf{P}_b, \mathbf{P}_{-b})}{P_b^{\text{total}}(\mathbf{P}_b, \mathbf{P}_{-b})} \leq Q_b^*, \quad \frac{R_b(\mathbf{P}_b^*, \mathbf{P}_{-b})}{P_b^{\text{total}}(\mathbf{P}_b^*, \mathbf{P}_{-b})} = Q_b^*$$

In other words, the optimal power control policy for the cell $b$: $\{\mathbf{P}_b^*, \mathbf{P}_{-b}\}$ is the optimal strategy for the original objective function. The proof of the converse issue of the Theorem 1 is completed.

The proposed energy-efficient power allocation is summarized in Algorithm 1. $l$ is the iteration index, $L$ denote the maximum number of iterations, and $\varsigma$ is the maximum tolerance.

---

**Algorithm 1**: Iterative Energy-efficient Power Allocation Algorithm

---

1: Initialize $l=0$, $\mathbf{P}_b(0) = \mathbf{0}$, and $\varsigma = 10^{-2}$.
2: For $b=1$ to $B$ do
3:    For $l=1$ to $L$ do
4:      Solve (21) for a given $Q_b(l-1)$ and obtain $\mathbf{P}_b(l)$.
5:      If $R_b(\mathbf{P}_b(l), \mathbf{P}_{-b}) - Q_b P_b^{\text{total}}(\mathbf{P}_b(l), \mathbf{P}_{-b}) \leq \varsigma$ and $P_{kb}(l) \geq p_{kb}, \forall k$.
6:       Then $\mathbf{P}_b^* = \mathbf{P}_b(l)$ and $Q_b^* = R_b(\mathbf{P}_b^*(l), \mathbf{P}_{-b}) / P_b^{\text{total}}(\mathbf{P}_b^*(l), \mathbf{P}_{-b})$.
7:       Break
8:      Else
9:       $Q_b(l) = R_b(\mathbf{P}_b(l), \mathbf{P}_{-b}) / P_b^{\text{total}}(\mathbf{P}_b(l), \mathbf{P}_{-b})$ and $l=l+1$.
10:    End if
11:   End For
12: End For

---

*C. Outer Beamformer Design with Subspace Tracking Algorithm*

One key challenge in supporting high quality and high rate multimedia service to moving users is getting accurate CSI quickly, but perfect CSI can never be known actually by receivers under moving users or fast time varying channel environment between cells. Therefore, it is important to develop adaptive outer beamformer tracking algorithm for MIMO networks.

Therefore, we develop an adaptive tacking algorithm for archiving the beamforming matrixes for the problem **P1.2**

$$LIF_b = \underbrace{\sum_{b' \neq b}^{B} \sum_{i=1}^{K} \left\| (\hat{\mathbf{H}}_{ib}^{b'} + \Delta \mathbf{H}_{ib}^{b'}) \mathbf{F}_b \right\|^2}_{\text{inter-cell interference leakage}} = \mathbf{F}_b^H \mathbf{\Phi}_b \mathbf{F}_b \quad (24)$$

where $\mathbf{\Phi}_b = \sum_{b' \neq b}^{B} \sum_{i=1}^{K} (\hat{\mathbf{H}}_{ib'}^b + \Delta \mathbf{H}_{ib'}^b)^H (\hat{\mathbf{H}}_{ib'}^b + \Delta \mathbf{H}_{ib'}^b) \mathbf{\Phi}_b = \sum_{b' \neq b}^{B} \sum_{i=1}^{K} \hat{\mathbf{H}}_{ib'}^{b\,H} \hat{\mathbf{H}}_{ib'}^b + (BK-1)\delta_e^2 \mathbf{I}_{d_k}$.

From (24), in order to compute the optimal outer beamforming vector $\mathbf{F}_b^*$ for the problem **P1.2** is to find the smallest eigenvalue corresponding to the smallest eigenvalues of the channel covariance matrix according the Theorem 1 in [10]. $\mathbf{\Phi}_b$ in an adaptive manager when minimize the total interference leakage. Note that $\mathbf{\Phi}_b$ is a Hermitian matrix where its eigenvectors are orthonormal. Therefore, based on the Courant–Fischer theorem, the optimal outer beamforming vector $\mathbf{F}_b^*$ to the smallest eigenvalues is obtained by minimizing the following Rayleigh quotient

$$\mathbf{F}_b^* = \arg \min_{\mathbf{F}_b} \frac{\mathbf{F}_b^H \mathbf{\Phi}_b \mathbf{F}_b}{\mathbf{F}_b^H \mathbf{F}_b} \quad (25)$$

Then, the Rayleigh quotient function for these Fourier transformation of the channel covariance matrix $\mathbf{\Phi}_b$ can be written as

$$J(\mathbf{F}_b) = -\frac{\mathbf{F}_b^H \mathbf{\Phi}_b \mathbf{F}_b}{\mathbf{F}_b^H \mathbf{F}_b} \quad (26)$$

The above Rayleigh quotient minimization issue can be handled adaptively by applying Newton method or gradient descent method. Among amount of descent methods, we adopt the conjugate gradient (CG) descent, which is suitable for Hermitian $\mathbf{\Phi}_b$, this does not require the matrix inversion applying the Newton method, and this scheme shows fast convergence.

In order to update $\mathbf{F}_b$, we can apply steepest gradient method to weight vector regarding the objective function $J(\mathbf{F}_b)$, which should be retracted to the Grassmann manifold. Then, the tangent vector on the above Grassmann manifold is written by

$$\mathbf{\Xi}_b = (\mathbf{I}_{d_k} - \mathbf{F}_b \mathbf{F}_b^H) \nabla J(\mathbf{F}_b) \quad (27)$$

Applying the above tangent vector, the solution achieved at each step of the above alternating minimization approach in (26), which aims to maximize the interference signal power leakage in (24), which is moved $\mathbf{F}_b$ along the following geodesic on the Grassmann manifold based on the steepest descent direction, namely

$$\mathbf{F}_b(\tau) = \mathbf{F}_b \mathbf{R} \cos(\mathbf{\Sigma} \tau) \mathbf{R}^H + \mathbf{\Lambda} \sin(\mathbf{\Sigma} \tau) \mathbf{R}^H \quad (28)$$

where $\mathbf{\Lambda} \mathbf{\Sigma} \tau \mathbf{R}^H$ is the compact singular decomposition of the tangent vector $\mathbf{\Xi}_b$ at the step size $\tau$.

However, maximizing $J(\mathbf{F}_b)$ along the geodesic given by $\nabla J(\mathbf{F}_b)$, may slow down the scheme convergence because of an alternation of those competing maximum-norm vectors from each iteration to each iteration. In order to reduce the complexity of the cost function, the conjugate search direction can be the combination of the old conjugate search direction as well as the new gradient, given by

$$\mathbf{\Theta}_b(l+1) = \mathbf{\Xi}_b(l) + \varpi(l) \tilde{\mathbf{\Theta}}_b(l) \quad (29)$$

$\varpi(l)$ can be got via the Polak Ribiére conjugacy formula [27],

which is the previous and new conjugate search direction, and it should be conjugate to the Hessian of $J(\mathbf{F}_b)$, given by

$$\varpi(l) = \frac{\text{tr}\left((\mathbf{\Xi}_b(l+1) - \tilde{\mathbf{\Xi}}_b(l))^H \mathbf{\Xi}_b(l+1)\right)}{\text{tr}(\mathbf{\Xi}_b^H(l)\mathbf{\Xi}_b(l))} \quad (30)$$

where $\tilde{\mathbf{\Xi}}_b(l)$ is the parallel translation $\mathbf{\Xi}_b(l)$ in the same way of $\mathbf{\Theta}_b(l)$. The parallel translation should keep all the directions within the conjugate tangent space at every iteration. The formula for obtaining $\Delta \mathbf{\Xi}_b(l)$ and $\tilde{\mathbf{\Xi}}_b(l)$ is

$$\begin{aligned}\tilde{\mathbf{\Theta}}_b(l) &= \left(\mathbf{F}_b(l)\mathbf{R}\sin(\mathbf{\Sigma}\tau) + \mathbf{\Lambda}\cos(\mathbf{\Sigma}\tau)\right)\mathbf{\Sigma}\mathbf{R}^H \\ \tilde{\mathbf{\Xi}}_b(l) &= \mathbf{\Xi}_b(l) - \left(\mathbf{F}_b(l)\mathbf{R}\sin(\mathbf{\Sigma}\tau) + \mathbf{\Lambda}(\mathbf{I} - \cos(\mathbf{\Sigma}\tau))\right)\mathbf{\Lambda}^H\mathbf{\Xi}_b(l)\end{aligned} \quad (31)$$

The design of outer beamforming matrices for the cell $b$ is concluded in the Algorithm 2.

---

**Algorithm 2**: IA based CGGM Algorithm for Outer Beamformer

---

1: Initialize $l=0$, $\tau$, $\varpi(l)$ and start with arbitrary beamforming matrix $\mathbf{F}_b(0)$, $\forall b=1,\cdots,B$ such that $\mathbf{F}_b^H(0)\mathbf{F}_b(0)=\mathbf{I}_{m_b}$. Set $\mathbf{\Xi}_b(0) = (\mathbf{I}_{d_b} - \mathbf{F}_b(0)\mathbf{F}_b^H(0))\nabla J(\mathbf{F}_b(0))$, $\tilde{\mathbf{\Xi}}_b(0) = \mathbf{\Xi}_b(0)$.
2: For $l=1,\cdots,L$ iteration
2.1: Compute the Rayleigh quotient function for each cell.
   $J(\mathbf{F}_b(l)) = -\mathbf{F}_b^H(l)\mathbf{\Phi}_b\mathbf{F}_b^H(l)/(\mathbf{F}_b^H(l)\mathbf{F}_b(l))$
2.2: Compute the compact decomposition of $\Delta\mathbf{\Xi}_b(l): \mathbf{\Lambda}\mathbf{\Sigma}\tau\mathbf{R}^H$.
2.3: Compute the geodesic on the Grassmann manifold
   $\mathbf{F}_b(\tau) = \mathbf{F}_b\mathbf{R}\cos(\mathbf{\Sigma}\tau)\mathbf{R}^H + \mathbf{\Lambda}\sin(\mathbf{\Sigma}\tau)\mathbf{R}^H$
2.4: Backtracking-Armijo step size ($\tau$) search
2.4.1: Given $\mathbf{F}_b(\tau)$ in a descended direction $\mathbf{\Theta}_b(n)$, set $\kappa \in (0, 0.5)$, $\upsilon > 1$
2.4.2: Backtracking: while
   $J(\mathbf{F}_b(\tau)) > J(\mathbf{F}_b(l)) + \kappa\tau\text{tr}((\nabla J(\mathbf{F}_b(l)))^T\mathbf{\Theta}_b(l))$, $\tau = \tau/\upsilon$.
2.4.3: Armijo: while $J(\mathbf{F}_b(\tau)) \leq J(\mathbf{F}_b(n)) + \kappa\tau\text{tr}((\nabla J(\mathbf{F}_b(n)))^T\mathbf{\Theta}_b(n))$
   and $J(\mathbf{F}_b(\upsilon\tau)) \leq J(\mathbf{F}_b(l)) + \kappa\upsilon\tau\text{tr}((\nabla J(\mathbf{F}_b(l)))^T\mathbf{\Theta}_b(l))$, $\tau = \upsilon\tau$.
2.5: Update the subspace $\mathbf{F}_b(l+1) = \mathbf{F}_b(\tau)$.
2.6: Compute $\tilde{\mathbf{\Theta}}_b(l+1)$ by projecting the gradient algorithm on the horizontal space at $\mathbf{F}_b(l+1)$
   $\mathbf{\Xi}_b = (\mathbf{I}_{d_k} - \mathbf{F}_b(l+1)\mathbf{F}_b^H(l+1))\nabla J(\mathbf{F}_b(l+1))$
2.7: Parallel transport the tangent vectors $\tilde{\mathbf{\Theta}}_b(l+1)$ and $\mathbf{\Xi}_b(l)$
   $\tilde{\mathbf{\Theta}}_b(l) = \left(\mathbf{F}_b(l)\mathbf{R}\sin(\mathbf{\Sigma}\tau) + \mathbf{\Lambda}\cos(\mathbf{\Sigma}\tau)\right)\mathbf{\Sigma}\mathbf{R}^H$
   $\tilde{\mathbf{\Xi}}_b(n) = \mathbf{\Xi}_b(l) - \left(\mathbf{F}_b(l)\mathbf{R}\sin(\mathbf{\Sigma}\tau) + \mathbf{\Lambda}(\mathbf{I} - \cos(\mathbf{\Sigma}\tau))\right)\mathbf{\Lambda}^H\mathbf{\Xi}_b(l)$
2.8: Compute the new tangent vector direction for next search as
   $\mathbf{\Theta}_b(l+1) = \mathbf{\Xi}_b(l) + \varpi(l)\tilde{\mathbf{\Theta}}_b(l)$
   where $\varpi(l) = \text{tr}((\mathbf{\Xi}_b(l+1) - \tilde{\mathbf{\Xi}}_b(l))^H\mathbf{\Xi}_b(l+1))/\text{tr}(\mathbf{\Xi}_b^H(l)\mathbf{\Xi}_b(l))$.
3: Set $l = l+1$, Repeat 2.3–2.9 until convergence.

---

For the above discussed CGGM Algorithm, the beamformer vectors $\mathbf{V}_{kb}^*$ and $\mathbf{F}_b$ should be updated at each time instant due to the time-varying channel, which leads to a high computational process when the mobile users or transmission and receive antennas is large. Differently from the CGGM algorithms, SM-based techniques apply a check block with a defined bound. According to (19) and (25), both the optimal inner and outer beamforming vectors are related to the channel covariance matrix, so we can set the defined bound based on the deviation between the previous channel covariance matrix and the current channel covariance matrix $\text{E}\{(\mathbf{\Phi}_{kb}^b(t) - \mathbf{\Phi}_{kb}^b(t-1))(\mathbf{\Phi}_{kb}^b(t) - \mathbf{\Phi}_{kb}^b(t-1))^H\}$ and $\text{E}\{(\mathbf{\Phi}_b(t) - \mathbf{\Phi}_b(t-1))\times(\mathbf{\Phi}_b(t) - \mathbf{\Phi}_b(t-1))^H\}$.

The SM technique usually has the following two steps: (1) information evaluation, and (2) parameter update. If the deviation between the previous and current channel covariance matrix are not changed largely, the bemaforming vector don't need to update frequently so that the step does not need much complexity, the system overall complexity will be reduced significantly.

Let $\mathcal{H}_b$ denote the set containing the updated outer beamforming vector $\mathbf{F}_b(t)$ at time instant $t$ is upper or down the bound given by $\Pi_b$ for the cell $b$, which is

$$\mathcal{H}_b = \begin{cases} \text{update } \mathbf{F}_b(t): \text{E}\{(\mathbf{\Phi}_b(t) - \mathbf{\Phi}_b(t-1))(\mathbf{\Phi}_b(t) - \mathbf{\Phi}_b(t-1))^H\} \geq \Pi_b \\ \mathbf{F}_b(t) = \mathbf{F}_b(t-1): \text{E}\{(\mathbf{\Phi}_b(t) - \mathbf{\Phi}_b(t-1))(\mathbf{\Phi}_b(t) - \mathbf{\Phi}_b(t-1))^H\} < \Pi_b \end{cases} \quad (32)$$

The above constrained SM adaptive strategy that will be given more details in the following subsection, when we apply it to increase the performance of the adaptive CGGM-based algorithm. The convergence as well as tracking performance will be enhanced due to the variable forgetting factor, whereas the system complexity can be reduced because of the beamforming vector-selective updates. Note that, due to the nature of dynamic wireless communication networks, the time-varying bound must be selected appropriately, in order to capture the characteristics of the dynamic networks environment and to improve the networks performance. The set-membership adaptive filtering framework for the CGGM IA adaptive algorithm can be concluded in Algorithm 3.

---

**Algorithm 3**: Proposed SM–CGGM IA Adaptive Algorithm

---

1: For each time instant $t = 1, \cdots, T$.
2: For each cell $b = 1, \cdots, B$.
3: Giving the bound $\Pi_b$.
4: Compute the deviation
   $\Delta_b(t) = \text{E}\{(\mathbf{\Phi}_b(t) - \mathbf{\Phi}_b(t-1))(\mathbf{\Phi}_b(t) - \mathbf{\Phi}_b(t-1))^H\}$
5: If $\Delta_b(t) \geq \Pi_b$
   Solve Rayleigh quotient (25) to obtain $\mathbf{F}_b(t)$ with conjugation gradient algorithm based Grassmann manifold by using **Algorithm 2**.
6: Else
   $\mathbf{F}_b(t) = \mathbf{F}_b(t-1)$
7: Repeat 3–6 until convergence.

---

## IV. STABLE FDPM-BASED SUBSPACE TRACKING FOR RECEIVE BEAMFORMER

Due to the fast varying time channel, robust and low computational complexity should be taken into account. We will develop a practical IA algorithm based on fast and robust fast data projection method (FDPM) subspace tracking algorithms to achieve the receive beamforming vector $\mathbf{U}_{kb}$ in this section.

## A. The Subspace Tracking

The subspace tracking tool plays a key role in the matrix optimization as well as signal processing, which we have already given some discussion in the subsection C of the Section III. The problem of the subspace tracking over the minimized the interference leakage can be analyzed as follows. The observation vectors at each mobile user' receiver at the instant time slot $t$ can be expressed as

$$\mathbf{x}_{kb}(t)=(\hat{\mathbf{H}}_{kb}^{b}+\Delta\mathbf{H}_{kb}^{b})\mathbf{F}_{b}\mathbf{V}_{kb}\mathbf{H}_{kb}^{SI}\mathbf{z}_{kb}+(\hat{\mathbf{H}}_{jb}^{b}+\Delta\mathbf{H}_{jb}^{b})\mathbf{F}_{b}\sum_{j=1,j\neq k}^{K}\mathbf{V}_{jb}\mathbf{s}_{jb} \\ +(\hat{\mathbf{H}}_{kb}^{b'}+\Delta\mathbf{H}_{kb}^{b'})\mathbf{F}_{b'}\sum_{b'\neq b}^{B}\sum_{i=1}^{K}\mathbf{V}_{ib'}\mathbf{s}_{ib'}+\mathbf{n}_{kb} \quad (33)$$

Note: in this process, the transmit beamforming vectors $\mathbf{F}_b$, $\mathbf{V}_{kb}$, $\mathbf{F}_{b'}$, and $\mathbf{V}_{ib'}$ are fixed. The observation covariance matrix can be given as obtained by $\mathbf{Q}_{kb}(t)=E\{\mathbf{x}_{kb}(t)\mathbf{x}_{kb}^H(t)\}$. The minor subspaces of $\mathbf{Q}_{kb}(t)$ are defined: the subspaces spanned by the minor eigenvectors of $\mathbf{Q}_{kb}(t)$ associated to the lowest eigenvalues.

Then, the goal of the developed algorithm is to iteratively update the receive beamforming vector $\mathbf{U}_{kb}(t)$ that minimizes the following Rayleigh quotient

$$\begin{aligned} &\underset{\mathbf{U}_{kb}(t)}{\text{Min}} \quad \mathbf{U}_{kb}^H(t)\mathbf{Q}_{kb}(t)\mathbf{U}_{kb}(t) \\ &\text{s.t.} \quad \mathbf{U}_{kb}^H(t)\mathbf{U}_{kb}(t)=\mathbf{I}_{d_k} \end{aligned} \quad (34)$$

It is easily observed that both the interference minimization problems in (33) and (34) are essentially the same objective form by minimizing the Rayleigh quotient. By tracking the minor subspaces of total interference covariance matrices $\mathbf{I}_{kb}(t)$ and $\mathbf{Q}_{kb}(t)$, we can obtain $\mathbf{U}_{kb}(t)$, respectively. A straightforward way to alive the subspace is to use an SVD on the covariance matrix $\mathbf{Q}_{kb}(t)$, which need $O(M^3)$ operations [28]. However, the FDPM algorithm requires low complexity requiring $O(d_{m_b}M)$ operations is utilized in minor subspace tracking.

## B. The Complex FDPM Algorithm

The FDPM inherits from the DPM, the overview of DPM scheme can be summarized below. The minimization of the objective function in (34) can be obtained iteratively applying the steepest descent approach. The orthonormal constraint in (34) is satisfied by applying the Gram–Schmidt orthonormalization iteratively. The DPM scheme is given by

$$\begin{aligned} \bar{\mathbf{x}}_{kb}(l) &= \mathbf{U}_{kb}^H(l-1)\mathbf{x}_{kb}(l) \\ \mathbf{T}_{kb}(l) &= \mathbf{U}_{kb}(l-1)+\alpha\mathbf{x}(l)\bar{\mathbf{x}}_{kb}^H(l) \\ \mathbf{U}_{kb}(l) &= \text{Gram-Schmid orthonormalization}\left(\mathbf{T}_{kb}(l)\right) \end{aligned} \quad (35)$$

where $\mathbf{U}_{kb}(l)$ means the tracked subspace at the $l$ th iteration and $\alpha$ is the step size parameter. Reversing the sign of $\alpha$, it yields a principal ($\alpha>0$) or minor ($\alpha<0$) tracking algorithm. If we user the Gram–Schmidt orthonormalization, the computational complexity of the above scheme is high. In order to reduce the complexity, the FDPM algorithm which constructs the following orthonormal matrix $\mathbf{A}_{kb}(l)$ to orthonormalize $\mathbf{T}_{kb}(l)$. This matrix $\mathbf{A}_{kb}(l)$ is defined as

$$\mathbf{A}_{kb}(l)=\mathbf{H}_{kb}^{\mathbf{a}}(l)\left(diag\{\mathbf{B}_{kb}^H(l)\mathbf{B}_{kb}(l)\}\right)^{-\frac{1}{2}} \quad (36)$$

where $\mathbf{B}_{kb}(l)=\mathbf{T}_{kb}(l)\mathbf{H}_{kb}^{\mathbf{a}}(l)$, and $\mathbf{H}_{kb}^{\mathbf{a}}(l)$ is a complex Householder matrix

$$\mathbf{H}_{kb}^{\mathbf{a}}(l)=\mathbf{I}_{d_k}-\frac{2}{\|\mathbf{a}(l)\|^2}\mathbf{a}(l)\mathbf{a}^H(l) \quad (37)$$

where $\mathbf{a}(l)=\bar{\mathbf{x}}_{kb}(l)-e^{j\theta}\|\bar{\mathbf{x}}_{kb}(l)\|\mathbf{e}$, the $\theta$ denotes the phase angle of the first element of $\bar{\mathbf{x}}_{kb}(l)$ and $\mathbf{e}=[1,0,\cdots,0]$. Consequently, $\mathbf{T}_{kb}(l)$ is orthonormalized by

$$\mathbf{U}_{kb}(l)=\mathbf{T}_{kb}(l)\mathbf{A}_{kb}(l)=\left(\mathbf{U}_{kb}(l-1)+\alpha\mathbf{x}(l)\bar{\mathbf{x}}_{kb}^H(l)\right)\mathbf{A}_{kb}(l) \quad (38)$$

In fact, the FDPM algorithm and other tracking schemes try to develop an orthonormalization matrix $\mathbf{A}_{kb}(l)$ to satisfy the following issue holding true

$$\mathbf{U}_{kb}^H(l)\mathbf{U}_{kb}(l)=\mathbf{A}_{kb}^H((l)\mathbf{T}_{kb}^H((l)\mathbf{T}_{kb}(l)\mathbf{A}_{kb}(l)=\mathbf{I}_{d_k} \quad (39)$$

where $\mathbf{T}_{kb}^H((l)\mathbf{T}_{kb}(l)=\mathbf{U}_{kb}^H(l-1)\mathbf{U}_{kb}(l-1)+\left(2\alpha+\alpha^2\|\mathbf{x}(l)\|^2\right)\bar{\mathbf{x}}(l)\bar{\mathbf{x}}_{kb}^H(l)$.

However, due to the time varying channel as well as the channel estimation uncertainty, it is hard to guarantee the product $\mathbf{U}_{kb}^H(l)\mathbf{U}_{kb}(l)=\mathbf{I}_{d_k}$ at each instant time slot. Therefore, it is important to develop a modified robust FDPM subspace tracking algorithm to achieve the receive beamforming vectors.

## C. Proposed Robust FDPM IA Algorithm

The FDPM subspace tracking IA algorithm transforms the IA problem in MIMO networks into an unconstrained IA problem, for a general multi-cell multi-user MIMO interference channel and eliminates the IUI as well as ICI among users while causing no interference to other serving users. The above complex FDPM-based minor subspace tracking algorithm is utilized to obtain IA through a training period, without any priori knowledge of those interference covariance matrices (8).

Assuming interference alignment is feasible [30], the interference alignment is achieved once the beamforming and receive combining vectors should satisfy

$$\begin{aligned} \mathbf{U}_{kb}^H(\hat{\mathbf{H}}_{jb}^{b}+\Delta\mathbf{H}_{jb}^{b})\mathbf{F}_b\mathbf{V}_{jb} &= 0, \quad \forall k, \forall b, \\ \mathbf{U}_{kb}^H(\hat{\mathbf{H}}_{kb}^{b'}+\Delta\mathbf{H}_{kb}^{b'})\mathbf{F}_{b'}\mathbf{V}_{ib'} &= 0, \quad \forall k, \forall b, \\ \mathbf{U}_{kb}^H(\hat{\mathbf{H}}_{kb}^{b'}+\Delta\mathbf{H}_{kb}^{b'})\mathbf{F}_b\mathbf{V}_{kb} &\neq 0, \quad \forall k, \forall b, \\ \text{rank}(\mathbf{U}_{kb}^H(\hat{\mathbf{H}}_{kb}^{b'}+\Delta\mathbf{H}_{kb}^{b'})\mathbf{F}_b\mathbf{V}_{kb}) &= \mathbf{I}_{d_k}, \quad \forall k, \forall b, \end{aligned} \quad (40)$$

Due to the effect of channel estimation uncertainty, In order to analyze the numerical stability, we focus on the deviation of $\mathbf{U}_{kb}^H(l)\mathbf{U}_{kb}(l)$ from the identity matrix. Towards this end, we define $\mathbf{U}_{kb}^H(l)\mathbf{U}_{kb}(l)=\mathbf{I}_{d_k}+\boldsymbol{\xi}_{d_k}$. Where $\boldsymbol{\xi}_{d_k}$ denote the deviation from the identity matrix $\mathbf{I}_{d_k}$. According to the literature [29], the fast and robust subspace tracking algorithm can be given as at the $l$ th iteration

$$\begin{aligned}
\bar{\mathbf{x}}_{kb}(l) &= \mathbf{U}_{kb}^H(l-1)\mathbf{x}_{kb}(l) \\
\mathbf{z}_{kb}(l) &= \mathbf{U}_{kb}(l-1)\bar{\mathbf{x}}_{kb}(l) \\
\mathbf{B}_{kb}(l) &= \mathbf{z}_{kb}(l)/\|\bar{\mathbf{x}}_{kb}(l)\| + \alpha \mathbf{x}(l)\|\bar{\mathbf{x}}_{kb}(l)\| \quad (41)\\
\mathbf{C}_{kb}(l) &= \mathbf{B}_{kb}(l)/\|\mathbf{B}_{kb}(l)\| - \mathbf{z}_{kb}(l)/\|\bar{\mathbf{x}}_{kb}(l)\| \\
\mathbf{U}_{kb}(l) &= \mathbf{U}_{kb}(l-1) + \mathbf{C}_{kb}(l)\bar{\mathbf{x}}_{kb}^H(l)/\|\bar{\mathbf{x}}_{kb}(l)\|
\end{aligned}$$

The above FDPM algorithm is numerically stable, more details of the proof can be seen in [15]. The design of receive beamforming matrices for all users is concluded in the Algorithm 4.

---

**Algorithm 4**: Stable FDPM IA Minor Subspace Tracking Algorithm

1: Initialization: Initialize random matrices $\mathbf{U}_{kb}(0)$, satisfy $\mathbf{U}_{kb}^H(0)\mathbf{U}_{kb}(0)=\mathbf{I}_{d_k}$ Initial step size: $\alpha_0 (\alpha_0 < 0)$
2: For $b=1$ to $B$ do
3:  For $k=1$ to $K$ do
4:   For $l=1$ to $L$ do
5:    For $b=1$ to $B$ do
     The cell $b$ transmit the signals for each user $k$: $\mathbf{F}_b(l)\mathbf{V}_{kb}(l)\mathbf{s}_{kb}(l), \forall k$. The transmit inner and outer beamforming vectors $\mathbf{V}_{kb}(l)$ and $\mathbf{F}_b(l)$ can be obtained based on (18), **Algorithm 1, Algorithm 2 and Algorithm 3**.
    End for
6:   The user $k$ in the cell $b$ receive the interference signals $\mathbf{x}_{kb}(l)$ according to (33).
    End for
7:   Compute the initialized vector $\mathbf{z}_{kb}(0)$, $\mathbf{B}_{kb}(0)$ and $\mathbf{C}_{kb}(0)$ according to (41).
8:   Determine step size: $\beta$. Backtracking-Armijo step size ($\alpha$) search.
8.1:  Given the obstruction vectors $\bar{\mathbf{x}}_{kb}(l-1)$ and $\mathbf{x}_{kb}(l-1)$
8.2:  Backtracking: while $\mathbf{U}_{kb}(l) > \mathbf{U}_{kb}(l-1) + \mathbf{C}_{kb}(l-1)\bar{\mathbf{x}}_{kb}^H(l-1)/\|\bar{\mathbf{x}}_{kb}(l-1)\|$, $\beta = \beta_0/\|\mathbf{x}_{kb}(l)\|$.
8.3:  Armijo: while $\mathbf{U}_{kb}(l) \leq \mathbf{U}_{kb}(l-1) + \mathbf{C}_{kb}(l-1)\bar{\mathbf{x}}_{kb}^H(l-1)/\|\bar{\mathbf{x}}_{kb}(l-1)\|$, $\beta = -\beta_0/\|\mathbf{x}_{kb}(l)\|$.
8.4:  Update the receive vector $\mathbf{U}_{kb}(l)$ based on (41).
   End for
  End for
9: Output: The postprocessing matrix $\mathbf{U}_{kb}(L)$ at the user $k$ in the cell $b$, $\forall k, \forall b$.

---

## V. SIMULATION RESULTS

In this session, we perform some simulations to evaluate the performance of our proposed algorithm and compare it with some existing algorithms. The compared existing two-tier beamforming solutions in multi-user multi-cell MIMO systems are following as:

(1) The scheme in [10] developed an outer beamformer with subspace alignments by minimizing the sum inter-group interference power from other groups and minus the weighted all desired group signal power, and applying a low complexity subspace tracking algorithm to achieve the beamformer by assuming perfect CSI.
(2) A two-stage beamforming framework based on SLNR was developed in [12], and the authors transformed the problem as a trace quotient problem, which aims to enhance significant sum rate gain with perfect CSI.
(3) Our proposed robust two-tier transmit and receive beamforming based interference alignment for multi-user multi-cell MIMO interference channels with considering channel error.

We set a cellular network with $B=3$ cells, and each of mobile user has two receive antennas. The channel vector for every mobile user is independently and the noise power is set as $\delta^2=1$. The circuit power at per antenna is $P_b^c = 30\text{dBm}$, and the basic power consumed at the BS is $P_b^o = 40\text{dBm}$. The moving velocity of each mobile user is denoted as $v$. The reciprocal efficiency of the power amplifier $\rho$ at per BS is 0.39.

### A. Performances Versus Transmit Power Constraint

Fig.2 compares the per cell throughput performance achieved by the three algorithms with transmit power constraint. Obviously, the algorithm [12] is better to all other algorithms. We can see that all the three algorithms ca achieve the same spectral efficiency in the small transmit power region, especially, the advantage of the method [12] I not obvious in the process. However, with the increase of transmit power, the method [12] outperforms the other two algorithms, and the effective SE of other two algorithms saturates to one certain level approximately, this is because the algorithm always try to enhance SLNR with higher power for SE maximization, while the our proposed algorithm prefers to reduces the interference power to others cells, as well as allocate the transmit power in order to enhance the overall energy efficiency so that it achieves more energy efficiency shown in Fig.3 than that in the algorithm [12].

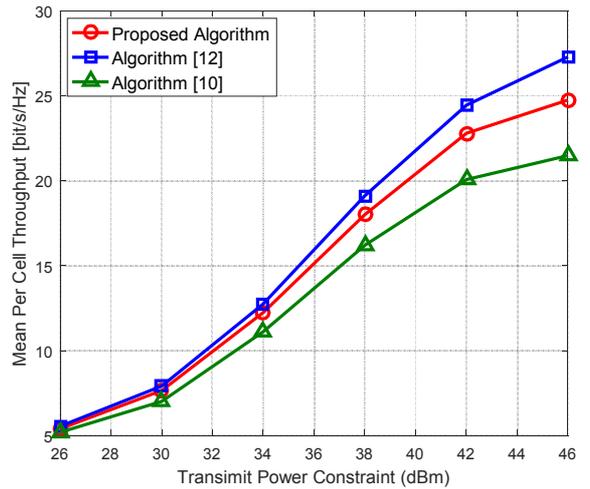

**FIGURE 2. The per cell throughput versus transmit power constraint when $K=4$, $M=8$, $\varepsilon=0.05$, and $v=5$ km/h.**

Fig. 3 demonstrates the EE performance of our proposed algorithm compared with other two algorithms under different transmit power constraint per BS. It is observed that the proposed algorithm outperforms other two methods in terms of EE. In the energy efficiency of the proposed algorithm obviously outperforms the other algorithms greatly at the middle-high transmission power region, namely, from 36 to 46dBm. This is because our proposed scheme considers energy-efficient power allocation and interference migration. Moreover, when the transmit power is more than 44dBm, the algorithm [10] is better to the algorithm [12]. Because in the algorithm [12], the SE gain fails to compensate for the negative effect of the energy consumption, leading to a lower energy efficiency. Numerical results also indicate that channel error causes the performance degradation for other two algorithms in terms of EE performance.

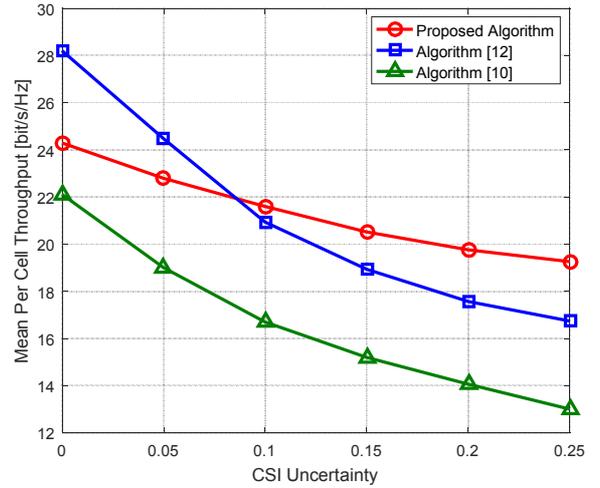

**FIGURE 4.** The per cell throughput against CSI uncertainty when $K=4$, $M=8$, $v=5$ km/h **and** $P_T=42$ dBm.

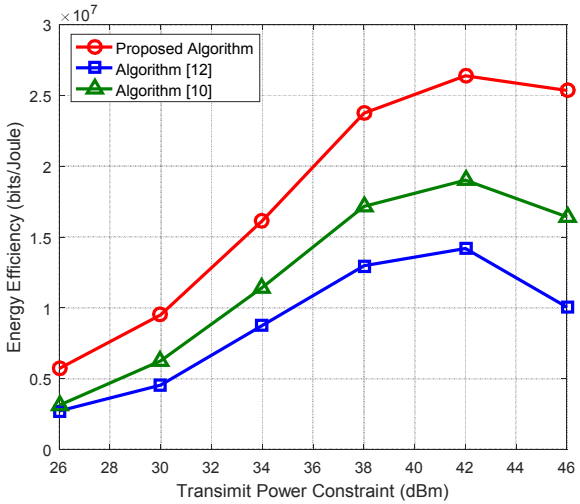

**FIGURE 3.** Energy efficiency versus transmit power constraint when $K=4$, $M=8$, $\varepsilon=0.05$, **and** $v=5$ km/h.

### B. Performance Against Channel Errors

Fig. 4 and Fig. 5 show the sum rate per cell and EE performance versus the channel estimation error for different schemes. We can observe that under the channel uncertainty environment, the proposed robust algorithm always obtains higher worst case energy efficiency than those of the other two algorithms, even the sum rate performance of per cell is little fewer than that in the algorithm [12]. Clearly, there is a tradeoff between the spectral efficiency and energy efficiency. In addition, the performance gap of the performance between our proposed scheme and other two solutions become bigger as the uncertainty degree increases, which illustrates the proposed scheme is more stable against the channel error. This is because of the foundation that with considering the channel estimation error, then both the two tier transmit and receive beamforming vectors are more robust against the channel uncertainty.

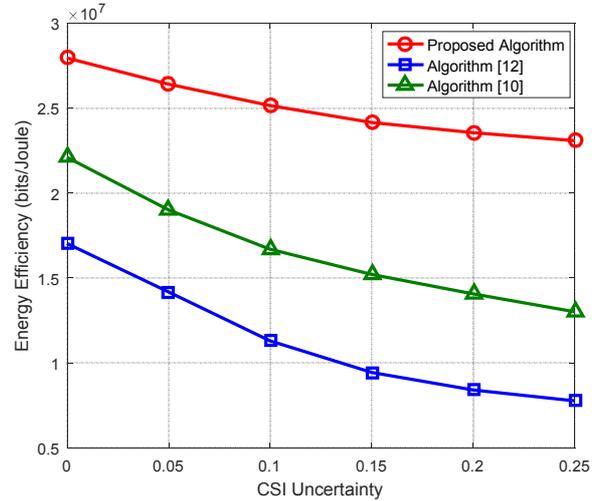

**FIGURE 5.** Energy efficiency against CSI uncertainty when $K=4$, $M=8$  $v=5$ km/h **and** $P_T=42$ dBm.

### C. Performance Versus Mobile Users' Velocity

Fig. 6 and Fig. 7 show the sum rate per cell and the EE performance versus the mobile user speed. The performance of the three schemes decrease when the mobile speed increases since higher velocity induces faster channel change. Especially, the performance of the algorithm [12] is more sensitive to the users speed, and its performance degrades significantly as 5ms the velocity become large. On the other hand, both our proposed scheme and the algorithm [10] are robust to mobile users' velocity. This is because both these two methods developed subspace tracking algorithms based on gradient method under time-varying channels. Moreover, the proposed algorithm performs best due to it apples low-complexity tracking scheme so that it is more suitable for the time varying channel environments.

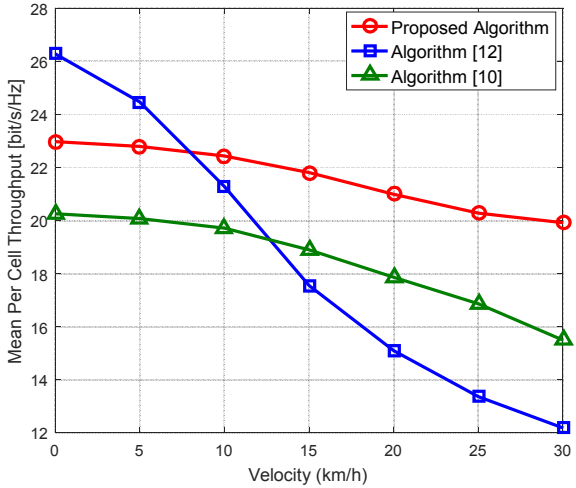

**FIGURE 6.** The per cell throughput versus the mobile user mobility under $K=4$, $M=8$, $P_T$=42dBm and $\varepsilon=0.05$.

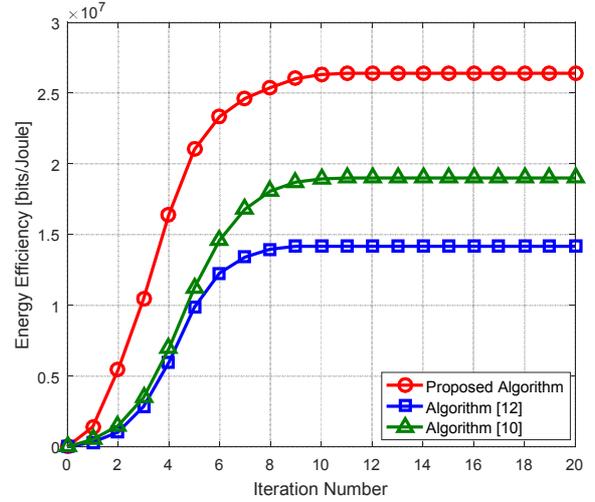

**Fig. 8.** The convergence of energy efficiency.

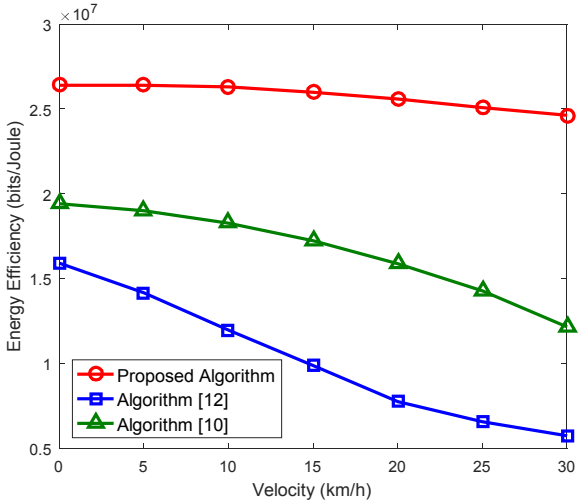

**FIGURE 7.** Energy efficiency versus the mobile user mobility under $K=4$, $M=8$, $P_T$=42 dBm and $\varepsilon=0.05$.

*E. Complexity*

Finally, we illustrate the convergence of the proposed algorithm and compare it with other schemes, when $K=4$, $M=8$, $\varepsilon=0.05$, $v=5$ km/h and $P_T$=42 dBm. From Fig. 10, it can be seen that the distributed algorithm iteratively converges to the final point about seven times, while other two schemes need the more number of converge iteration, about nigh times in [12] and eleven times in [10]. This is because we apply low complexity subspace tracking algorithms to achieve transmit and receive beamformer with considering the channel estimation errors.

The computational complexity for the three are analyzed in this subsection. For our proposed scheme, the computational complexity of inter bemaforming vector and energy efficient power is $O(K^2M^2d_k+BK^2M^2d_kL)$. The computational complexity of the set-membership adaptive filtering framework is $O(K(2d_k+(2d_k^2+8d_k+6)))$. For the outer beamformer, the modified conjugate gradient Grassmann manifold subspace tracking algorithm needs $O(Kd_k(d_k+1)M^2L)$ computational operations. In this case, the training period of stable FDPM-ST IA algorithm needs $O((M+N)Kd_kL)$ operations for achieving the outer beamformer. To specially mention, the total computational complexity of our proposed algorithm is related to $O(K^2M^2d_k(B+1)+K(2d_k+(2d_k^2+8d_k+6))+Kd_k(d_k+1)M^2L+(M+N)Kd_kL)$. The computational complexity of the two-tier precoding strategy in [10] requires about $O(2Kd_bM^2L)$ operations. Two-stage beamformer design in [12] should need about $O(2Kd_bM^2NL)$ computational operations for the iterative training period. Through the above computational complexity analysis for the three approaches, we can observe that our proposed robust transceiver design needs a little more complexity than that of other two approaches, but it is an acceptable complexity in multi-user multi-cell networks. In addition, the proposed scheme considers more conditions to improve the system performance, which has better performance than that of other two methods.

## VI. CONCLUSION

In this paper, we design a robust transceiver design by exploiting the IUI and ICI based on interference alignment for multi-user multi-cell MIMO interference networks with channel uncertainties. The two-tier beamforming scheme tries to minimize the interference leakage to other cells or users with considering energy efficient power allocation. In particular, to decrease the computational complexity for the outer beamforming, we develop a modified set-membership conjugate gradient Grassmann manifold subspace tracking algorithm to achieve the outer beamforming matrix efficiently. Then, we propose a practical interference alignment based on fast and robust fast data projection method (FDPM) subspace tracking algorithm, to achieve the robust receive beamformer under channel uncertainty. Numerical results show that our proposed robust transceiver design achieves better performance compared with some existing methods in terms of the sum rate and the energy efficiency.

REFERENCES

[1] G. Zheng, I. Krikidis, C. Masouros, S. Timotheou, D. A. Toumpakaris and Z. G. Ding, "Rethinking the role of interference in wireless networks," *IEEE Comm. Mag.*, vol.52, no.11, pp.152–158, Nov. 2014.
[2] S. Timotheou, G. Zheng, C. Masouros and I. Krikidis, "Exploiting constructive interference for simultaneous wireless information and power transfer in multiuser downlink systems," *IEEE J. Sel. Areas Comm.*, vol. 34, no. 5, pp. 1772-1784, May 2016.
[3] S. M. Razavi, "Unitary Beamformer Designs for MIMO Interference Broadcast Channels," *IEEE Trans. Signal Processing*, vol. 64, no. 8, pp. 2090-2102, Apr. 2016.
[4] S. W He, Y. G. Huang, S. Jin and L. X. Yang, "Energy efficient coordinated beamforming design in multi-cell multicast networks," *IEEE Comm. Let.*, vol.19, no.6, pp.985–988, June 2015.
[5] G. Nauryzbayev, and E. Alsusa, "Enhanced multiplexing gain using interference alignment cancellation in multi-cell MIMO networks," *IEEE Trans. Inform. Theory*, vol. 62, no. 1, pp. 357-369, Jan. 2016.
[6] M. J. Kim, H. H. Lee, and Y. C. Ko, "Limited feedback design for interference alignment on two-cell interfering MIMO-MAC," *IEEE Trans. Veh. Technol.*, vol. 64, no. 9, pp. 4019-4030, Sept. 2015.
[7] J. Tang, D. K. C. So, E. Alsusa, K. A. Hamdi and A. Shojaeifard, "Energy efficiency optimization with interference alignment in multi-cell MIMO interfering broadcast channels," *IEEE Trans. Comm.*, vol. 63, no. 7, pp. 2486-2499, July 2015.
[8] E. Björnson, M. Bengtsson and B. Ottersten, "Optimal multiuser transmit beamforming: a difficult problem with a simple solution structure," *IEEE Signal Process. Mag.*, vol. 31, no. 4, pp. 142-148, July 2014.
[9] O. E. Ayach, S. Rajagopal, S. Abu-Surra, Z. Pi, and R. W. Heath, Jr., "Spatially sparse precoding in millimeter wave MIMO systems," *IEEE Trans. Wireless Comm.*, vol. 13, no. 3, pp. 1499–1513, Mar. 2014.
[10] J. Chen and V. Lau, "Two-tier precoding for FDD multi-cell massive MIMO time-varying interference networks," *IEEE J. Sel. Areas Comm.*, vol. 32, no. 6, pp. 1230–1238, Jun. 2014.
[11] J. Nam, A. Adhikary, J. Ahn, and G. Caire, "Joint spatial division and multiplexing: Opportunistic beamforming, user grouping and simplified downlink scheduling," *IEEE J. Sel. Topics Signal Process.*, vol. 8, no. 5, pp. 876–890, Oct. 2014.
[12] D. G. Kim; G. Lee and Y. C. Sung, "Two-stage beamformer design for massive MIMO downlink by trace quotient formulation," *IEEE Trans. Comm.*, vol.63, no.6, pp.2200–2211, June 2015.
[13] P. Nagaradjane, and T. Muthu, "Performance of cooperative multi-cell downlink communication aided by polarization-multiplexing under limited feedback constraints," *IEEE Access*, vol. 4, no. 8, pp. 3479-3488, 2016.
[14] Y. Ren, Y. Wang, C. Q, and Y. Liu, "Multiple-beam selection with grouping-based interference alignment with IA-cell for hybrid beamforming in massive MIMO systems," *IEEE Access*, vol.PP, no.99, pp.1-10, 2017.
[15] Q. Yang, T. Jiang, C. Jiang, Z. Han and Z. Zhou, "Joint optimization of user grouping and transmitter connection on multi-Cell SNR blind interference alignment," *IEEE Access*, vol. 4, no.1, pp. 4974-4988, 2016.
[16] C. Masouros and G. Zheng, "Exploiting known interference as green signal power for downlink beamforming optimization," *IEEE Trans. Signal Processing*, vol.63, no.14, pp.3628–3640, Jul., 2015.
[17] P. Cao, A. Zappone and E. A. Jorswieck, "Grouping-based interference alignment with IA-cell assignment in multi-cell MIMO MAC under limited feedback," *IEEE Trans. Signal Processing*, vol. 64, no. 5, pp. 1336-1351, March1, 2016.
[18] N. Zhao, F. R. Yu, H. Sun and M. Li, "Adaptive power allocation schemes for spectrum sharing in interference-alignment-based cognitive radio networks," *IEEE Trans. Veh. Technol.*, vol. 65, no. 5, pp. 3700-3714, May 2016.
[19] J. Park, Y. C. Sung, D. G. Kim and H. V. Poor, "Outage probability and outage-based robust beamforming for MIMO interference channels with imperfect channel state information," *IEEE Trans. Wireless Comm.*, vol.11, no.10, pp.3561–3573, Oct. 2012.
[20] H. D. Nguyen, R. Zhang and H. T. Hui, "Multi-cell random beamforming: achievable rate and degrees of freedom region," *IEEE Trans. Signal Processing*, vol.61, no.14, pp.3532–3544, Jul. 2013.
[21] R. F. Guiazon, K.-K. Wong and D. Wisely, "Capacity analysis of interference alignment with bounded CSI uncertainty," *IEEE Wireless Comm. Let.*, vol.3, no.5, pp.505–508, Oct. 2014.
[22] J. Yoon, W. Y. Shin and H. S. Lee, "Energy-efficient opportunistic interference alignment," *IEEE Comm. Let.*, vol.18, no.18, pp.30–33, Jan. 2014.
[23] K. Baddour and N. Beaulieu, "Autoregressive models for fading channel simulation," in *Proc. IEEE GLOBECOM*, 2001, vol. 2, pp. 1187–1192.
[24] L. Zhang, Y. C. Liang, Y. Xin and H. V. Poor, "Robust cognitive beamforming with partial channel state information," *IEEE Trans. Wireless Comm.*, vol. 8, no. 8, pp. 4143 –4153, Aug. 2009.
[25] A. Wiesel, Y. Eldar, and S. Shamai, "Linear precoding via conic optimization for fixed MIMO receivers," *IEEE Trans. Signal Processing*, vol. 54, no. 1, pp. 161–176, Jun. 2006.
[26] Y. Wu, J. Wang, L. Qian, and R. Schober, "Optimal power control for energy efficient D2D communication and its distributed implementation," *IEEE Commun. Lett.*, vol.1, no.1, pp.1–4, Feb. 2015.
[27] A. Edelman, T. A. Arias and S. T. Smith, "The geometry of algorithms with orthogonality constraints," *SIAM J. Matrix Anal. Appl.*, vol. 20, no. 2, pp. 303–353, 1998.
[28] X. G. Doukopoulos and G. V. Moustakides, "The fast data projection method for stable subspace tracking," in *13th Europ. Signal Process. Conf.*, Sep. 2005.
[29] C. Yetis, T. Gou, S. Jafar and A. Kayran, "On feasibility of interference alignment in MIMO interference networks," *IEEE Trans. Signal Process.*, vol. 58, no. 9, pp. 4771–4782, Sept. 2010.